\documentclass{ws-mpla}

\newcommand{\beeq}{\begin{equation}}
\newcommand{\eneq}{\end{equation}}
\newcommand{\beeqa}{\begin{eqnarray}}
\newcommand{\eneqa}{\end{eqnarray}}

\begin{document}

\markboth{Alessandra Feo}
{Predictions and recent results in susy on the lattice}

%
\catchline{}{}{}{}{}
%

\title{
{\vspace{-1.2em} \parbox{\hsize}{\hbox to \hsize 
{\hss \normalsize UPRF-2004-25}}} \\
PREDICTIONS AND RECENT RESULTS IN SUSY ON THE LATTICE}

\author{\footnotesize ALESSANDRA FEO}

\address{Dipartimento di Fisica, Universit\`a di Parma and INFN Gruppo Collegato di Parma,\\
Parco Area delle Scienze, 7/A. 43100 Parma,
Italy \\
feo@fis.unipr.it}

\maketitle


\begin{abstract}
In this brief review, I summarize the current theoretical knowledge in supersymmetry on the lattice,
with special emphasis on recent results in the framework of $N=1$ supersymmetric Yang Mills theory,
Wess-Zumino model and Yang-Mills theory with extended supersymmetries. 

\keywords{Supersymmetry; Lattice field theory; Super Yang-Mills.}
\end{abstract}

\ccode{PACS Nos.: include PACS Nos.}

\section{Introduction}	
Non-perturbative studies of supersymmetric gauge theories exhibit many fascinating properties 
\cite{seiberg} which are of great physical interest
\footnote{See for example, \cite{intriligator,ferretti} for recent reviews on supersymmetric gauge 
theories and related topics \cite{bagger,weinberg}.}.
For this reason, much effort has been dedicating to formulating 
lattice version of supersymmetric theories. See for example \cite{kaplan,feo} for recent reviews
on the subject at the latest Lattice conferences.

While much is known analytically, the hope is that the lattice  
would provide further information and confirm the existing analytical calculations.
The lattice formulation has been successful to extract non-perturbative dynamics in field theory,
specially in QCD, and may be able to provide additional information also for supersymmetry. 
Four dimensional supersymmetric gauge theories are good laboratories for non supersymmetric QCD and 
its extensions. 
Whether supersymmetry is or not an exact symmetry is a question that must be settle by going beyond 
perturbation theory. 

This letter is organized as follows. In Section 2, non-perturbative aspect of the four dimensional 
super Yang-Mills are reviewed. In Section 3, lattice results of this model using 
Wilson fermions are described while in Section 4 results using chiral fermions are presented.
Recent examples of exact supersymmetry on the lattice including the Wess-Zumino model and final remarks are 
given in Section 5.

\section{Non perturbative effects in $N=1$ super Yang-Mills theory}
The dynamics of $N=1$ super Yang-Mills is similar to QCD: confinement of the colored degrees of freedom 
and spontaneous chiral symmetry breaking. Its ground state consists of at least 
$N_c$ different vacua parametrized by the imaginary phase of a non zero gluino condensate
\cite{witten,snvz} related by discrete $Z_{2 N_c}$ transformations on the gluino fields.
Once one of the $N_c$ vacua is chosen, the $Z_{2 N_c}$ symmetry group spontaneously breaks down to 
the $Z_2$ group. 
One also expects that, in each of those ground states, the spectrum of the model
consists of colorless bound states of gluinos and gluons (see \cite{armoni,strassler} for 
an interesting relation between QCD and super Yang-Mills). 
The discrete chiral symmetry breaking is expected to be broken by a non-zero gluino condensate 
while the confinement is realized by colorless bound states described by the effective action
belonging to chiral supermultiplets. 
Moreover, non perturbative effects may cause a supersymmetry anomaly, as has been discussed in 
\cite{shamir}. In this case, not only the mass term would be responsible for a soft breaking.
Only a study of the continuum limit of the lattice supersymmetric Ward-Takahashi
identity can give us a better understanding on this subject. Some step forward in this direction
can be found in \cite{taniguchi,feo2}. 

The continuum action density 
for $N=1$ super Yang-Mills theory and a gauge group $SU(N_c)$ with a vector boson $A_\mu$ and 
a 4-component Majorana spinor $\lambda^a$ reads
\beeq
{\cal L} = -\frac{1}{4} F_{\mu \nu}^a(x) \, F_{\mu \nu}^a(x) + \frac{1}{2} \, \bar \lambda^a(x) \gamma_\mu 
({\cal D}_\mu \lambda(x))^a \, ,
\label{action}
\eneq
where the  Majorana spinor satisfies the Majorana condition $\bar \lambda^a = {{\lambda}^a}^T C$, and
$D_\mu$ is the covariant derivative in the adjoint representation,
${\cal D}_\mu \lambda^a = \partial_\mu \lambda^a + g f_{abc} A_\mu^b \lambda^c$.

This density action~(\ref{action}) is invariant under the continuum supersymmetric transformations,
\beeq
\delta A_\mu(x) = -2 g \bar\lambda(x) \gamma_\mu \varepsilon \, ,  \quad 
\delta \lambda(x) = -\frac{i}{g}\sigma_{\rho\tau} F_{\rho\tau}(x) \varepsilon \, , \quad
\delta \bar\lambda(x) = \frac{i}{g} \bar \varepsilon \sigma_{\rho\tau}F_{\rho\tau}(x) \, ,
\label{transform}
\eneq
where $\sigma_{\rho\tau} = \frac{i}{2} [\gamma_\rho,\gamma_\tau] $, $\lambda = \lambda^a T^a$ 
and $\varepsilon $ is a global Grassmann parameter with Majorana properties.
For $N=1$ super Yang-Mills theory the associated Noether current, $S_\mu$, reads
\beeq
S_\mu(x) = - F_{\rho \tau}^a(x) \sigma_{\rho \tau} \gamma_\mu \lambda^a(x) \, .
\eneq
This current is conserved, $\partial_\mu S_\mu = 0$ (if the fields satisfy the equations of motion) 
and satisfies the relation $\gamma_\mu S_\mu(x) = 0$. 

The density action (\ref{action}) has the global $U(1)_\lambda$ chiral symmetry
\beeq
\lambda \to e^{-i \varphi \gamma_5} \lambda \, , \qquad \qquad 
\bar \lambda \to \bar \lambda e^{-i \varphi \gamma_5} \, .
\label{a6}
\eneq
This symmetry is anomalous because the divergence of the axial current, 
$J^5_\mu = \bar \lambda \gamma_\mu \gamma_5 \lambda$, is non zero, 
\beeq
\partial_\mu J^5_\mu = \frac{N_c g^2}{32 \pi^2} \varepsilon^{\mu \nu \rho \sigma} F_{\mu \nu}^a 
F_{\rho \sigma}^a \, .
\eneq
The anomaly leaves a $Z_{2 N_c}$ subgroup of $U(1)_\lambda$ unbroken. 
Introducing of a non-zero gluino mass term in Eq.~(\ref{action}), 
${\cal L}_{mass} = m_{\tilde g}\bar \lambda^a \lambda^a$, breaks supersymmetry softly
(which implies that the non-renormalization theorem and cancellation
of divergences are preserved \cite{girardello}).
In the supersymmetric case, $m_{\tilde g}=0$, the $U(1)_\lambda$ symmetry is unbroken if
$\varphi \equiv \frac{k \pi}{N_c}$ for $(k=0,1,\cdots,2 N_c -1)$; here $\varphi$ is defined so that,
$\Theta_{SYM} \to \Theta_{SYM} - 2 N_c \varphi$.
$Z_{2N_c}$ is expected to be spontaneously broken to $Z_2$ by a value of 
$\big< \bar \lambda \lambda \big> \not = 0$ \cite{witten}. The consequence of this 
spontaneous chiral symmetry breaking is the existence of a first order phase
transition at $m_{\tilde g} = 0$. That means the existence of $N_c$ degenerate ground
states with different orientations of the gluino condensate $(k=0,\cdots,N_c -1)$,
\beeq
\big<\bar \lambda \lambda \big> = c \Lambda^3 e^{\frac{2 \pi i k}{N_c}} \, ,
\label{a2}
\eneq
where $\Lambda$ is the dynamical scale of the theory which can be calculated on the lattice, for instance, 
while $c$ is a numerical constant which depends on the renormalization scheme 
used to compute $\Lambda$.
Eq.~(\ref{a2}) shows the dependence on the gauge group.  
For $SU(2)$ two degenerate ground states with opposite sign of the
gluino condensate, $\big< \bar \lambda \lambda \big> < 0$ and 
$\big< \bar \lambda \lambda \big> > 0$, appear \cite{kirchner}, while for $SU(3)$ there 
are three degenerate vacua at $k =k_c$ (for a first numerical study for $SU(3)$ see \cite{su3}). 

The value of the gluino condensate (\ref{a2}), for the gauge group $SU(2)$, has been calculated in the eighties 
by using two different methods. One is based on strong coupling instanton calculations,\cite{strong},
while the second one is based on weak coupling instanton calculations, \cite{weak}. They 
give different result for $c$ (in Eq.~(\ref{a2})) and this was known as the $\frac{4}{5}$ puzzle.  
Various discussion about the validity of both methods can be found in the literature \cite{amati}.
More recently, a third elegant method \cite{davies} calculates the gluino condensate directly in the 
semiclassical approximation. This method gives results in agreement with the weak coupling instanton 
approximation \cite{weak,finnell} and confirm the correctness of the weak coupling instanton 
result. 

\subsection{Light hadron spectrum}
Effective lagrangians are extremely useful to describe strongly interacting theories in terms of 
their relevant degrees of freedom. 
The effective lagrangian for $N=1$ super Yang-Mills, also known as supersymmetric gluodynamics,  
which describes interactions between gluons and gluinos, was found by Veneziano and Yankielowicz 
(VY) \cite{veneziano}.
In terms of a minimal number of degrees of freedom contained in the $S$,
\beeq
S = \frac{3}{32 \pi^2 N_c} \mbox{Tr} W^2 \, ,
\eneq 
where $W_\alpha$ is the supersymmetric field strength, the VY lagrangian reads 
\beeq
{\cal L}_{VY}= \frac{9 N_c^2}{4 \alpha} \int d^2 \theta d^2 \bar \theta (S^\dagger S)^{1/3} + 
\frac{N_c}{3} \int d^2 \theta \bigg[S \mbox{log}(\frac{S}{\Lambda^3})^{N_c} - N_c S) \bigg] + h.c. \, .
\label{vy}
\eneq
where $\Lambda$ is a renormalization group invariant scale related with the super Yang-Mills parameter.
The K\"ahler term here is ambiguous. The one presented in Eq.~(\ref{vy}) is the simplest one compatible with
the symmetries of the theory. 

The chiral superfield $S$ at the component level is written as \cite{ferrara}
\beeq
S(y) = \phi(y) + \sqrt{2} \theta \chi(y) + \theta^2 F(y) \, ,
\eneq
where $\phi$ represent the scalar and pseudoscalar gluinoballs while $\chi$ is their fermionic partner.
However, no physical glueballs appear in the VY effective lagrangian.
Some attempts in order to include glueballs in the VY lagrangian have already appeared in the literature 
\cite{farrar,farrar2,cerdeno} (while a completely different approach in \cite{minkowski} 
claims dynamical breaking of supersymmetry and its absence from the spectrum).
Although it is tempting to say that $F$ represent the scalar and pseudoscalar glueballs, it is
an auxiliary field. Hence these states are not represented in the VY lagrangian. 
It has been stressed in \cite{sannino} that the VY lagrangian is not an effective 
lagrangian in the same sense, as for example, the pion chiral lagrangian, which describes 
light degrees of freedom and can, therefore, be systematically improved by introducing 
high derivative terms. 
Generalization of the VY lagrangian with more degrees of freedom are discussed in the literature
\cite{farrar,farrar2,cerdeno}. These results are in substantial agreement with 
old lattice simulations using Wilson fermions \cite{montvay,evans}, but they were still away from the 
supersymmetric limit.

Recently, Sannino and Shifman \cite{sannino} constructed an effective lagrangian of the VY type
for two non-supersymmetric theories which are orientifold daughters of supersymmetric gluodynamics
and at large $N_c$ they recover the bosonic sector of the VY action constructed for super Yang-Mills.
The spectrum consists of a massive pseudoscalar (eta prime) and the associated scalar fields. 
At large $N_c$ these states are mapped in the pseudoscalar and scalar super Yang-Mills gluinoball.
In Ref. \cite{merlatti}, the VY lagrangian has been extended to incorporate 
supersymmetric glueballs states ($R=0$) while and in Ref. \cite{fms}, using the extended VY lagrangian of 
Ref. \cite{merlatti} together with QCD results, it was finally deduced that gluinoballs are 
the lightest states in super Yang-Mills. 

\section{Lattice formulation of supersymmetric gauge theories} 
The problem of putting a supersymmetric theory on the lattice has been addressed in the past
by several authors \cite{curci,dondi,old,golterman}. 
One obstacle arises from the fact that the supersymmetry algebra is actually an extension 
of the Poincar\'e algebra, which is explicitly broken on the lattice. Schematically
one has, $\{Q,\bar Q  \} = 2 P$. We know that Poincar\'e invariance is achieved automatically 
in the continuum limit without fine tuning since operators that violate Poincar\'e invariance are all irrelevant.
On the other hand, if the supersymmetry theory contains scalar fields one can have scalar mass terms 
that break supersymmetry. Since these operators are relevant, fine tuning is necessary 
in order to cancel their contributions.

Another problem is the question of how to balance bosonic and fermionic modes, the numbers
of which are constrained by the supersymmetry: the naive lattice fermion 
formulation results in the doubling problem \cite{nielsen},
and produces a wrong number of fermions. The problem can be treated as in QCD
by using different fermion formulations. Let us briefly summarize 
those which have applications in supersymmetric theories. In the first of them, one ends up
with breaking of chiral symmetry. This is the Wilson formulation.

\subsection{$N=1$ super Yang-Mills theory Wilson fermions}
Wilson fermions are undoubled, non-chiral and ultra-local.
In the Wilson formulation of Curci and Veneziano (CV) \cite{curci}, it is proposed to construct a non supersymmetric
discretized $N=1$ super Yang-Mills with a supersymmetric continuum limit. 
Here, supersymmetry is broken by the lattice itself, by the Wilson term and a soft breaking due 
to the gluino mass is present. Supersymmetry is recovered in the continuum limit by tuning the bare parameters $g$ 
and the gluino mass $m_{\tilde{g}}$ (through the hopping parameter) to the supersymmetric limit.
The supersymmetric (and chiral) limits are both recovered simultaneously at $m_{\tilde{g}}=0$ 
(see \cite{montvay} for a review on $N=1$ super Yang-Mills using Wilson fermions). 

The CV effective action suitable for Monte Carlo simulations is
\beeq
S_{CV} = \beta \sum_{pl} \bigg( 1 - \frac{1}{N_c} \mbox{Re} \mbox{Tr} U_{\mu \nu} \bigg) - \frac{1}{2} 
\mbox{log} \, \mbox{det} Q[U] \, ,
\eneq
where the bare coupling is given by $\beta \equiv 2 N_c/g^2$ for the gauge group $SU(N_c)$.
The fermion matrix $Q$ for the gluino is defined by
\beeq
\hspace{-0.6 cm} Q_{yb,xa}[U] \equiv 
  \delta_{xy} \delta_{ab} - k \sum_{\mu=1}^4 \big[ \delta_{y, x + \hat \mu} (1 + \gamma_\mu) V_{ba,x \mu} + 
   \delta_{y + \hat \mu, x} (1 - \gamma_\mu) V^T_{ba,x \mu} \big] \, ,
\label{a15}
\eneq
where $k$ is the hopping parameter defined as $k = 1/(2(4 + m_0 a))$, $m_0$ is the bare mass,
and the matrix for the gauge field link in the adjoint representation is
\beeq
V_{ab,x \mu} \equiv  V_{ab,x \mu}[U] \equiv 
\frac{1}{2} \mbox{Tr}( U^\dagger_{x \mu} T_a U_{x \mu} T_b ) \, .  
\label{a9}
\eneq
The fermion matrix for the gluino in Eq.~(\ref{a15}) is not hermitian but it satisfies the relation
$Q^\dagger = \gamma_5 Q \gamma_5$. That allows for the definition of the hermitian
fermion matrix $\tilde Q \equiv \gamma_5 Q$. 
The path integral over the Majorana fermions gives the Pfaffian,
\beeqa
\int [d \lambda] e^{-\frac{1}{2} \bar \lambda Q \lambda} =
\int [d \lambda] e^{-\frac{1}{2} \lambda^T C Q \lambda} = Pf(M) \, ,
\eneqa
where $M \equiv C Q$ is an antisymmetric matrix.

It is easy to see that $Pf(M) = \pm\sqrt{det \, Q}$. In the effective CV action the absolute
value of the Pfaffian is taken into account (this may cause the sign problem). 
The spectral flow is a method who checks the value of the sign of the Pfaffian. 
Results of Refs.\cite{campos,farchioni} show that below the critical line 
$k_c(\beta)$, corresponding to zero gluino mass $(m_{\tilde{g}} =0)$, negative Pfaffians 
practically never appear. The method of simulation with non-even number of flavors is 
based on the multi-bosonic algorithm proposed by L\"uscher \cite{luscher} where a 
two-step variant using a noisy correction step \cite{kennedy},
has been developed by Montvay in \cite{montvay2,montvay3} called the two-step 
multibosonic (TSMB) algorithm.
In the two-step variant, to represent the fermion determinant one uses a first polynomial 
${\cal P}_{n_1}^{(1)}(x)$ for a crude approximation realizing a fine correction by another polynomial
${\cal P}_{n_2}^{(2)}(x)$ that satisfies, 
${\displaystyle \lim_{n_2 \to \infty}} {\cal P}_{n_1}^{(1)}(x) {\cal P}_{n_2}^{(2)}(x) = x^{-N_f/2}$,
for $x \in \, [\varepsilon,\lambda]$.
The fermion determinant is approximated as \cite{montvay2}
\beeqa
\mbox{det}(Q^\dagger Q)^{N_f} \simeq \frac{1}{\mbox{det} \, P_{n_1}^{(1)}(Q^\dagger Q) 
\mbox{det} \, P_{n_2}^{(2)}(Q^\dagger Q)} \, .
\eneqa
Unquenched results for the gauge group $SU(2)$ using TSMB have been reported in 
Refs.\cite{kirchner,campos,farchioni,peetzproc} and more recently in \cite{peetz}, 
while for $SU(3)$ a preliminary results is in Ref.\cite{su3}.
Previous quenched simulations can be found in \cite{koutsoumbas,hernandez,arroyo,arroyo2}.

By studying the pattern of chiral symmetry breaking, 
through the study of the first order phase transition of the gluino condensate it is then 
possible to determine the value of the critical 
hopping parameter which correspond to the supersymmetric limit ($m_{\tilde{g}}=0$). 
In \cite{kirchner,su3}, for a fixed value of 
$\beta$, it is introduced a gluino mass term that breaks supersymmetry and then it is tuned in order 
recover supersymmetry in the continuum limit. 
At the supersymmetric (chiral) limit, a first order phase transition (or a crossover) should shows up as a 
two double peak structure in the distribution of some order parameter (the gluino condensate, in this case), 
indicating that the corresponding 
$k_c$ is the critical hopping parameter corresponding to the supersymmetric limit. 
By increasing the volume the tunneling between the two ground states becomes less and less 
probable and at some point practically impossible. Outside the phase transition region,
the observed distribution can be fitted by a single Gaussian but in the transition region 
a good fit can only be obtained with two Gaussians. 

How do we know we are restoring supersymmetry in the continuum limit?
\begin{itemize}
\item
This can be achieved for example, by investigating the low-lying mass spectrum and comparing with theoretical 
predictions. In \cite{campos,peetzproc}, simulations near the value of $k_c$, have been performed. 
An accurate study of this issue is non-trivial not only from the computational point of view but also due to 
several different theoretical formulations. An independent method for checking the supersymmetry restoration 
would be demanding. Recently, in \cite{peetz} it is shown that the pseudoscalar gluinoball is indeed the
lighter particle of the supermultiplet at the value of the gluino mass measured. The latter findings seem
to be more consistent with the recent predictions of Ref. \cite{fms}.

\item
Another independent way to study the supersymmetry restoration in the Wilson formulation 
is through the study of the supersymmetric Ward-Takahashi identity (WTi).
This has been achieved both numerically \cite{farchioni} and in lattice perturbation theory \cite{taniguchi,feo2}.
Let us briefly summarize these results.
\end{itemize}

\subsection{The supersymmetric Ward-Takahashi identity}
Numerical simulations of the WTi \cite{farchioni} and more recently in \cite{peetz}
have been performed in order to determine 
non-perturbatively a substracted gluino mass and the mixing coefficients of the supercurrent.
The supersymmetric WTi in a numerical simulation reads \cite{farchioni},
\beeq
\big< {\cal O}(y) \nabla_\mu S_\mu(x) \big> +
Z_T Z_S^{-1} \big< {\cal O}(y)\nabla_\mu T_\mu(x) \big> = m_R Z_S^{-1} \big< {\cal O}(y) \chi(x) \big> \, ,
\label{numerical}
\eneq
and can be computed at fixed $\beta$ and $k$. By choosing two elements of the $4 \times 4$ matrices,
having previously chosen the operator insertion ${\cal O}(y)$ in Eq.~(\ref{numerical}), 
a system of two equations can be solved for $Z_T Z_S^{-1}$ and $m_R Z_S^{-1}$ \cite{farchioni}.
The results in Ref. \cite{farchioni} show a restoration of supersymmetry in the continuum 
limit up to $O(a)$ effects. 
The vanishing gluino mass, extrapolated when determine $m_R Z_S^{-1}$, occurs at a value 
of the hopping parameter 
in agreement to the one calculated using the pattern of chiral symmetry breaking \cite{kirchner}. 

The supersymmetric WTi has been also studied in lattice perturbation theory, up to one loop order, 
in two different papers, \cite{taniguchi,feo2}.
In Ref. \cite{taniguchi} the bare WTi is written and from here the axial and supersymmetric
WTi are determined. 
Taniguchi shows \cite{taniguchi} that the additive mass correction appearing in the supersymmetric
WTi coincide with that from the axial $U(1)_R$ symmetry, as suggested by Curci and Veneziano in 
\cite{curci}. 
On the other hand, in \cite{feo2}, it is written the renormalized supersymmetric WTi, 
\beeqa
&& Z_S \big< O \nabla_\mu S_\mu(x) \big> + Z_T \big< O \nabla_\mu T_\mu(x) \big> - 
2 (m_0 - \tilde{m}) Z_{\chi}^{-1} \big< O \chi^R(x) \big> + \nonumber \\
&& Z_{CT} \big< \frac{\delta O} {\delta \bar \xi(x)}|_{\xi = 0} \big> - 
Z_{GF} \big< O \, \frac{\delta S_{GF}}{\delta \bar \xi(x)}|_{\xi = 0} \big> -
Z_{FP} \big< O \, \frac{\delta S_{FP}}{\delta \bar \xi(x)}|_{\xi = 0} \big> \nonumber \\ 
&& + \sum_j Z_{B_j} \big< O B_j \big>=0 \, ,
\label{renorm}
\eneqa
and the coefficient $Z_T$ is calculated in lattice perturbation theory at one loop order, in 
the off-shell regime. $Z_T$ is in good agreement with the one determined in \cite{farchioni}.
In \cite{feo2} it is also shown that the structure of the supercurrent mixing involves a more 
complicated structure that the one proposed by Curci and Veneziano \cite{curci} and in 
\cite{taniguchi}.

\section{Ginsparg-Wilson fermions}
A key element for the construction of a chiral lattice theory, i.e., a theory in which chiral 
and flavor symmetries can be preserved on the lattice, without fermion doubling, is the choice of 
a lattice Dirac operator $D$, that satisfies the Ginsparg-Wilson (GW) relation \cite{ginsparg}
\beeq
\gamma_5 D + D \gamma_5 = a D \gamma_5 D 
\label{gw}
\eneq
and the hermiticity condition $D^{\dagger} = \gamma_5 D \gamma_5$. 
The operator should also be local, gauge covariant and have a number of further properties
\cite{luscher2,luscher3,hasenfratz,laliena,hasenfratz2,neuberger,neuberger2}.

An explicit expression for this operator has been founded by Neuberger \cite{neuberger,neuberger2}
\beeq
D = \frac{1}{a} \bigg( 1 - X \, (X^\dagger X)^{-1/2} \bigg) \, , \qquad \qquad X = 1 - a D_w \, ,
\label{D}
\eneq
\noindent 
where
\beeq
D_w = \frac{1}{2} \gamma_\mu ( \nabla^\star_\mu + \nabla_\mu ) - \frac{a}{2} \nabla^\star_\mu \nabla_\mu 
\label{Dw}
\eneq
and 
\beeq
\nabla_\mu \phi(x) = \frac{1}{a}(U_\mu(x) \phi(x + a \hat \mu) - \phi(x)) \, , \,  
\nabla_\mu^\star \phi(x) = \frac{1}{a}(\phi(x) - U_\mu^\dagger(x - a \hat \mu) \phi(x - a \hat \mu))   
\eneq
are the forward and backward lattice derivative, respectively. Either overlap fermions or domain wall
fermions, satisfy the GW relation. Let us briefly review applications of domain wall fermions 
to super Yang-Mills.

\subsection{$N=1$ with domain wall fermions}
$N=1$ super Yang-Mills has been also studied on the lattice using the domain wall fermion (DWF) approach. 
Application of DWF in supersymmetric theories has been explored in \cite{neuberger3,kaplan2}
and also suggested in \cite{nishimura}, with a different approach as \cite{kaplan2}.
First Monte Carlo simulations for $N=1$ $SU(2)$ super Yang-Mills with DWF, using the lines of 
Refs. \cite{neuberger3,kaplan2} are in \cite{fleming}. This formulation, even at non-zero 
lattice spacing does not require fine-tuning. The condensate is non-zero even for small volume and 
small supersymmetry breaking mass where zero mode effects due to gauge fields with fractional topological
charge appear to play a role. 

DWF were introduced in \cite{kaplan22} and were further developed in \cite{narayanan,shamir2}.
For a recent review on DWF for supersymmetric gauge theories see \cite{vranas2}. 
The DWF approach is defined by extending the space-time
to five dimensions. Also a non-zero five dimensional mass or domain wall height $m_0$, which 
controls the number of flavors, is present. 
The size of the fifth dimension, $L_s$, and free boundary conditions for the fermions are 
implemented. As a result, the two chiral components of the Dirac fermion are separated with 
one chirality bound exponentially on one wall and the other on the opposite wall.
For any value of $a$ the two chiralities mix only by an amount that decreases exponentially 
as $L_s \to \infty$. For $L_s = \infty$, the chiral symmetry is expected to be exact even 
at finite lattice spacing. Therefore, there is no need for fine tuning.
DWF offer the opportunity to separate the continuum limit, $a \to 0$, from the chiral limit, $L_s \to \infty$. 
For the time being, only the study of the gluino condensate has been performed, while the spectrum 
of the theory was not possible to measure on the small lattices used in \cite{fleming}.
In any case, due to the improved chiral properties of DWF, even though they appear to be 
much more expensive than non-chiral fermions, it would be nice to have more results with DWF or
overlap fermions, for example, concerning the spectrum of the theory.

\section{Exact lattice supersymmetry: where and how}
As we have already discussed, improving lattice supersymmetry is rather difficult for gauge theories.
In fact, most supersymmetric theories, as for example $N=2$ or $N=4$ super Yang-Mills, 
contain scalar bosons which typically produce supersymmetry violating relevant operators, 
which need to be fined tuned in some way.
For $N=1$, as we have seen, only a fine tuning is needed in order to eliminate the mass term.
Unfortunately, because there is no discrete version of supersymmetry which can be implemented 
to forbid scalar masses and unwanted relevant operators,
it is desirable to construct lattice structures which directly display at least
a subset of exact supersymmetry in order to decrease the number of fine tuning to do.
This idea has been applied to the two and four dimensional Wess-Zumino model and to   
super Yang-Mills with extended supersymmetries. Let us summarize some of them.

\subsection{The lattice Wess-Zumino Model}
The Wess-Zumino model \cite{wess} has been extensively studied on the lattice 
both, numerically and in lattice perturbation theory. 
Several interesting achievements can be found in the old literature,
starting from \cite{dondi} where a lattice regularized Wess-Zumino theory was constructed 
which although non-local, exhibited an explicit supersymmetry. Other old attemps can be 
found in \cite{old}, where also the case of the four dimensional 
lattice Wess-Zumino model is included. 
An interesting old reference is \cite{golterman} where a successful construction of 
the two dimensional lattice Wess-Zumino model using Wilson fermions was realized. 
Also the Ward identities resulting from a 
lattice generalization of the continuum supersymmetry hold for each order in perturbation theory
\cite{golterman}. 

More recently, Fujikawa and Ishibashi have studied in detail lattice supersymmetry for the 
four dimensional $N=1$ Wess-Zumino model. In \cite{fujikawa} a lattice regularization of the supersymmetric 
Wess-Zumino model is studied by using the Ginsparg-Wilson operator 
\footnote{A previous attempt that uses a RG transformation can be found in \cite{bietenholz} and then in \cite{so}.}.
Ref. \cite{fujikawa} pointed out a 
conflict between the lattice chiral symmetry and the Majorana condition for Yukawa couplings. 
Also, three different examples of lagrangian for the Dirac fermions with Yukawa couplings are shown. 
The first one is the most natural one consistent with lattice chiral symmetry which is softly broken by the 
mass term. The second possible lagrangian incorporates continuum chiral symmetry with Yukawa coupling which is
softly broken by the mass term but not explicitly lattice chiral, even for $m=0$. Yet, another 
lagrangian which is suggested by the analysis of lattice chiral gauge symmetry.

In \cite{fujikawa2} further investigations on the arguments raised in \cite{fujikawa} are performed. 
In particular it is shown that the conflict between lattice chiral symmetry 
and the Majorana condition by the presence of Yukawa couplings already noted in \cite{fujikawa} 
is related in an essential way to the basic properties of the Ginsparg-Wilson operator, 
namely, locality and absence species doubling. 
In \cite{fujikawa3} it is pointed out that the major obstacle for the construction of a lattice supersymmetric 
theory is the failure of the Leibniz rule, as has been previously noticed by \cite{dondi}.
In \cite{fujikawa} it is suggested that a lattice version of a perturbatively finite theory
preserves supersymmetry to all orders in perturbation theory, in the sense that the supersymmetry breaking terms
induced by the failure of the Leibniz rule become irrelevant in the continuum limit. Differences between 
Ginsparg-Wilson fermions and Wilson fermions are also analyzed. 

$N=2$ Wess-Zumino model has been investigated in \cite{fujikawa4} defined on the $d=2$ Euclidean lattice in 
connection with the restoration of the Leibniz rule in the limit $a \to 0$.
In \cite{fujikawa4} also the differences between Wilson and Ginsparg-Wilson fermions are investigated and 
the effects of extra interactions introduced by an analysis of a Nicolai mapping.
As for the Wilson fermions, it induces a linear divergence to individual tadpole diagrams in the 
limit $a \to 0$, which is absent when Ginsparg-Wilson fermions are used.
The $N=2$ Wess-Zumino model in \cite{fujikawa4} also investigates the effects of the extra couplings
introduced by an analysis of Nicolai mapping. 

The Wess-Zumino model has also been studied by Catterall and collaborators in a series of interesting 
papers. In \cite{catterall}, a lattice version for the two dimensional Wess-Zumino model with
$N=2$ supersymmetry is presented. The lattice prescription chosen has the merit of preserving exactly a 
single supersymmetric invariance at finite lattice spacing.
From the form of the transformations they have derived a set of Ward identities which would be 
satisfied in the continuum limit. In \cite{catterall} it is argued that the presence of one 
exact symmetry (together with the finiteness of the continuum theory) guarantees that the full symmetry 
is restored without fine tuning in the continuum limit. These claims have been checked 
by an explicit numerical simulation of the Euclidean lattice theory and using a Fourier Hybrid Monte Carlo 
algorithm \cite{catterall2} to handle the fermionic integration of dynamical fermions in order 
to check the equality of the mass gaps in the boson and fermion sectors and to check the 
first non trivial lattice Ward identity. 

Further numerical investigations of the two dimensional Wess-Zumino model can be found in \cite{catterall3} 
using the action analyzed by Golterman and Petcher \cite{golterman} where a perturbative proof was given that 
the continuum supersymmetric Ward identities are recovered without fine tuning in the limit of vanishing 
lattice spacing by using Wilson fermions. The numerical simulations in \cite{catterall3} demonstrate
the existence of important non-perturbative effects in finite volumes which modify these conclusions:
It appears that in certain region of parameter space the vacuum state can contain solitons corresponding
to soliton configurations which interpolate between different classical vacua. 
In the background of these solitons supersymmetry is partially broken and a light fermion mode is observed. 
At fixed coupling the critical mass separating phases of broken and unbroken supersymmetry 
appears to be volume dependent. Ref. \cite{catterall3} 
also discussed the implication on supersymmetry breaking.

A very interesting paper is \cite{catterall4} where 
it is known that certain theories with extended supersymmetry can be discretized in such a way as to preserve 
an exact fermionic symmetry. In the simplest model of this kind, this residual supersymmetric 
invariance is actually a BRST symmetry. As an example, the supersymmetric 
quantum mechanics which possesses two such BRST symmetries is investigated and there it is shown that
at the quantum level, the continuum BRST symmetry is preserved in the lattice theory.
Similar conclusions are reached for the two-dimensional complex Wess-Zumino model and imply 
that all the supersymmetric Ward identities are satisfied exactly on the lattice. 
In \cite{catterall4} several numerical results supporting these conclusions are presented.
More recently, in \cite{catterall5}, it is studied how to  
construct lattice sigma models in one, two and four dimensions which exhibit an 
exact fermionic symmetry. These models are discretized and twisted versions of conventional 
supersymmetric sigma models with $N=2$ supersymmetry are showed. As an example, the $O(3)$ 
supersymmetric sigma model in two dimensions is presented.

In a recent work by D'Adda, Kanamori, Kawamoto and Nagata \cite{dadda} 
a new formulation which realizes exact twisted supersymmetry for all the supercharges 
on a lattice by twisted superspace formalism is proposed. This is achieved by introducing a 
mild lattice noncommutativity to preserve Leibniz rule on the lattice. 
Explicit examples of $N=2$ twisted supersymmetry invariant BF and Wess-Zumino models in two dimensions
are shown.

Other examples on the construction of lattice non-gauge supersymmetric models up to four 
supercharges in various dimensions have been studied by Giedt and Poppitz \cite{giedt2}.
Here it is shown the conditions under which the interacting lattice theory can exactly 
preserve one or more nilpotent anticommuting supersymmetries written in the superfield formalism.
In some cases, one exact supersymmetry guarantees the recovering of the continuum limit 
without fine tuning.

Wipf and collaborators \cite{wipf} investigated a class of two dimensional Wess-Zumino models
by using the nonlocal and antisymmetric SLAC derivative. They show that SLAC derivatives
allow chiral fermions without doublers and also minimizes supersymmetry breaking lattice
artifacts. In \cite{wipf},
the supercharges of the lattice Wess-Zumino models are obtained by dimensional reduction of Dirac 
operators in high-dimensional spaces. The normalizable zero modes of the models with $N=1$ and $N=2$ 
supersymmetry are counted and constructed in the weak and strong-coupling limits.

In a recent work, \cite{bonini}, a lattice formulation of the four dimensional Wess-Zumino model 
that uses Ginsparg-Wilson fermions and keeps exact supersymmetry to the full action is presented. 
The supersymmetry transformation that leaves 
the action invariant at finite lattice spacing is determined by performing an iterative procedure in the 
coupling constant. The closure of the algebra, generated by this transformation is also showed.
In \cite{bonini} a simple lattice Ward identity up to order $O(g)$ is verified. 

Ref. \cite{campostrini} contains a careful writeup of the study of dynamical
supersymmetry breaking by non perturbative lattice techniques in a class of the two-dimensional $N=1$ 
Wess-Zumino models using the Hamiltonian formalism and analyze the phase diagram of a couple 
of simple models based on cubic or quadratic prepotential by explicit numerical simulations with Green 
Function Monte Carlo methods \cite{linden}.
The results for the model with cubic prepotential indicate unbroken
supersymmetry while for quadratic prepotentials the existence of two phases of broken and unbroken supersymmetry are
showed. 
The idea has been previously applied by Beccaria and Rampino, \cite{beccaria2}, 
where it is studied supersymmetry breaking 
in the lattice $N=1$ Wess-Zumino model by the world-line path integral algorithm. 
The ground state energy and supersymmetric Ward identities are exploited to support 
the expected symmetry breaking 
in finite volume. Non-Gaussian fluctuations of the topological charge are discussed and related to the infinite 
volume transition. In \cite{beccaria} the lattice $N=1$ Wess-Zumino model in two dimensions is studied
by constructing a sequence $\rho^{(L)}$ of exact lower bounds on its ground state energy density 
$\rho$, converging to $\rho$ in the limit 
$L\to\infty$. The bounds $\rho^{(L)}$ can be computed numerically on a finite lattice with $L$ sites 
and can be exploited to discuss dynamical symmetry breaking. The transition point is determined and 
compared with previously and independent method results from Campostrini and collaborators, 
\cite{campostrini}, with good agreement.
High precision study of the structure of $d=4$ supersymmetric Yang-Mills quantum mechanics is 
showed by Wosiek and Campostrini in \cite{wosiek,wosiek2,wosiek3}.

\subsection{Super Yang-Mills theory}
In Ref.~\cite{itoh}, Itoh, Kato, Sawanaka, So and Ukita, presented an entirely new approach towards 
a realization of super Yang-Mills theory on the lattice. 
The action consists of staggered fermions \cite{kogut} and the plaquette 
variables distributed in the Euclidean space with a particular pattern. 
The system is shown to have fermionic symmetries relating the fermion and the link variables.
The gauge action has a novel structure. Though it is the ordinary plaquette action,
two different couplings are assigned in the ``Ichimatsu pattern'' or the checkered pattern. In the 
naive continuum limit, the fermionic symmetry survives as a continuum (or an $O(a^0)$) symmetry. 
The transformation of the fermion is proportional to the field strength multiplied by the difference 
of the two gauge couplings in this limit. 
Ref.~\cite{suzuki} examines compatibility and difficulties on how 
to accomodate Majorana and Weyl fermions in various dimensional Euclidean lattice gauge theories.

In a series of interesting papers, Kaplan and collaborators presented a new approach to constructing 
lattices for gauge theories with extended supersymmetry \cite{kaplan3,kaplan4,kaplan5}.
The lattice theories themselves respect certain supersymmetries, which in many cases allows the 
target theory to be obtained in the continuum limit without fine tuning. 
This method exploits an orbifold construction \cite{arkani} for spatial lattices 
in Minkowski space (and then for Euclidean space), and can be generalized to more complicated theories 
with additional supersymmetry and more spacetime dimensions 
(see \cite{kaplan} for a detailed review on this subject). A challenging issue would be to dynamical simulate
these theories. However, there is apparently a sign problem 
\cite{giedt} which may be resolved due to some encouraging results \cite{giedt3}.

Another formulation of super Yang-Mills theories with extended supersymmetry on hypercubic lattices of various 
dimensions keeping one or two supercharges exactly has been realized by Sugino, 
see \cite{sugino,sugino2} and references therein. 
It is interesting to mention a related topic. Non-perturbative supersymmetric theories have been also 
studied using the supersymmetric discrete light cone quantization (SDLCQ) formulation, 
see for example papers by Pinsky and collaborators, \cite{pinsky,pinsky2,pinsky3}, with very interesting 
numerical results. 
The SDLCQ formulation of the transverse lattice does not have species doubling \cite{pinsky4}. 

\section*{Acknowledgments}
I am grateful to M. Beccaria, M. Bonini, W. Bietenholz, M. Campostrini, G. F. De Angelis, P. Merlatti, 
H. Panagopoulos, F. Sannino for enlightening discussions.
It is also a pleasure to thank R. Brower, S. Catterall, P. De Forcrand, S. Ghadab, J. Giedt, M. Golterman, 
I. Kanamori, N. Kawamoto, K. Nagata, M. Shifman, H. So, Y. Taniguchi, P. Van Baal, A. Wipf, J. Wosiek, 
for useful discussions.

\end{document}